\documentclass[aip,preprint]{revtex4-1}
\usepackage{graphicx}
\usepackage{float}
\usepackage{color}

\begin{document}


\title{Engineered spin-valve type magnetoresistance in Fe$_3$O$_4$-CoFe$_2$O$_4$ core-shell nanoparticles} 

\author{P. Anil Kumar}
\affiliation{Centre for Advanced Materials, Indian Association for the Cultivation of Science, Kolkata, 700032, India}
\affiliation{Department of Engineering Sciences, Uppsala University, P.O. Box 534, SE-751 21 Uppsala, Sweden}
\author{Sugata Ray}
\affiliation{Centre for Advanced Materials, Indian Association for the Cultivation of Science, Kolkata, 700032, India}
\affiliation{Department of Materials Science, Indian Association for the Cultivation of Science, Kolkata, 700032, India}
\author{S. Chakraverty}
\affiliation{Centre for Advanced Materials, Indian Association for the Cultivation of Science, Kolkata, 700032, India}
\affiliation{RIKEN Center for Emergent Matter Science (CEMS), Wako 351-0198, Japan}
\author{D. D. Sarma}
\email[]{Also at Jawaharlal Nehru Centre for Advanced Scientific Research, Bangalore. Electronic mail: sarma@sscu.iisc.ernet.in}
\affiliation{Centre for Advanced Materials, Indian Association for the Cultivation of Science, Kolkata, 700032, India}
\affiliation{Solid State and Structural Chemistry Unit, Indian Institute of Science, Bangalore, 560012, India}
\affiliation{Department of Physics and Astronomy, Uppsala University, Box - 516, 75120 Uppsala, Sweden}
\affiliation{Council for Scientific and Industrial Research - Network of Institutes for Solar Energy (CSIR-NISE), New Delhi, India}

\date{\today}

\begin{abstract}
Naturally occurring spin-valve-type magnetoresistance (SVMR), recently observed in Sr$_2$FeMoO$_6$ samples, is suggestive of the possibility of decoupling the maximal resistance from the coercivity of the magnetic metallic grains, thereby providing an additional handle to tune the \textit{MR} of a material for technological advantages. In this paper, we present the first evidence that this can indeed be achieved in specifically designed and fabricated core-shell nanoparticle systems, realized here in terms of Fe$_3$O$_4$-CoFe$_2$O$_4$ core-shell nanocrystals. Here, the soft magnetic  Fe$_3$O$_4$ nanocrystals form the core, with the hard magnetic and highly insulating CoFe$_2$O$_4$ as the shell, providing a magnetically switchable tunnel barrier that controls the magnetoresistance of the system, instead of the magnetic properties of the magnetic grain material, Fe$_3$O$_4$, and thus proving the feasibility of engineered SVMR structures.
\end{abstract}


\maketitle 

The tunneling magnetoresistance (TMR) phenomenon \cite{Julliere54225} became technologically attractive after the realization of large magnetoresistance values. \cite{Moodera743273} In most of the naturally occurring polycrystalline TMR materials such as Fe$_3$O$_4$, CrO$_2$, La$_{0.67}$Sr$_{0.33}$MnO$_{3}$,\cite{Coey803815,Coey72734,Li711124,Hwang772041} the grain boundaries are found to act as the insulating, non-magnetic barrier. An important characteristic of such TMR systems is that the magnetoresistance, \textit{MR}, has a peak at the magnetic coercivity ($H_{C}$) of the grain material.\cite{Coey803815,Coey72734,Li711124} As a consequence, the \textit{MR} response of the system to an external magnetic field cannot be, in general, varied independent of the coercive field of the ferromagnetic metallic grains. However, it would be desirable to have this flexibility of choosing the zero of the \textit{MR} independent of the magnetic properties of the grain material, thereby providing us with an additional handle to design applications of such materials. For such manipulations of \textit{MR} as a function of $H$ we have recently shown \cite{Anil100262407} that the concept of ``dipolar biasing'' can be used very efficiently; however, this approach works well only in the very low magnetic field regime due to the intrinsic low strength of dipolar interactions compared to exchange interactions. An alternate approach suggests itself based on a different type of \textit{MR} reported in Sr$_2$FeMoO$_6$ (SFMO) \cite{Sarma98157205} where the \textit{MR} was shown to peak at a magnetic field higher than the magnetic coercivity of the material. This, and several other qualitative deviations of the experimentally observed behavior from the case of the usual TMR systems were shown\cite{Sarma98157205,Jana22346004,Ray9447007} to arise from the magnetic nature of the insulating barrier layer. Essentially, the departure of the zero \textit{MR} state from the coercive field was suggested to be controlled by the grain boundary material that was believed to be a hard magnetic insulator acting like a spin valve. This novel TMR mechanism has been termed as spin-valve-type magnetoresistance (SVMR)\cite{Sarma98157205} and its most essential ingredient is a magnetic insulating barrier with a coercive field higher than the coercive field of the metallic magnetic grains. It is to be noted that few known examples of SVMR relied on the grain boundary material fulfilling this criterion of being a harder magnet not by choice, but by accident. If these ideas are correct, it suggests a route to manipulate the electron tunneling process between the magnetic, metallic grains by deliberately creating an insulating, magnetic barrier whose spin orientation relative to that of the metallic grain can be controlled by the application of a suitable magnetic field. The relative orientation of the grain and the grain boundary spins will be determined by the relative strengths of the magnetic coercivities of the two parts. \par In order to engineer a TMR structure with a pronounced SVMR behavior, a convenient choice would be a core-shell nanomaterial where the core is made of a soft magnetic material with a substantial spin polarization and the shell should be a magnetic insulator with a high magnetic anisotropy and coercivity.  Additionally, the spin filtering property of CoFe$_2$O$_4$ \cite{Ramos91122107,Matzen101042409,Ramos78180402} makes it a very suitable choice for the barrier layer. We find that a system composed of Fe$_3$O$_4$ (FO) core and CoFe$_2$O$_4$ (CFO) shell suits the above criteria very well. Spinel Fe$_3$O$_4$ is a well known soft magnetic material with a coercivity of few Oe\cite{Ozdemir141351} in the bulk form and few hundred Oe for nanoparticles\cite{Caruntu405801} with a high degree of spin polarization.\cite{Liu833531,Fonin43487} Fe$_3$O$_4$ polycrystalline, thin film and nanocrystal samples have been studied\cite{Liu833531,Li837049,Anil20127} for its tunneling magnetoresistance property. CoFe$_2$O$_4$ also forms in spinel crystal structure and is a highly insulating magnetic material with a large coercivity, in contrast to a non-magnetic barrier.\cite{Chinnasamy832862} Spin filtering property of CFO has already been reported. \cite{Ramos91122107,Matzen101042409,Ramos78180402} It allows one to use non-magnetic metallic electrodes in MTJ systems, thereby widening the choice of material combinations for the study and application of tunneling magnetoresistance phenomenon. This choice also presents an added advantage due to the near perfect lattice matching between the core and the shell materials, allowing almost epitaxial growth of the shell material, CoFe$_2$O$_4$, over the core, Fe$_3$O$_4$. We chose a colloidal chemistry method to prepare the core-shell nanoparticles. This method is reported to produce uniform sized and high quality magnetic nanoparticles of both ordinary\cite{Park3891} and core-shell\cite{Masala81015} types. Our magnetization and magnetoresistance measurements on this specifically designed core-shell TMR structure establishes the influence of the magnetic barrier on the magnetoresistance of a TMR structure unambiguously.

\par Fe$_3$O$_4$ nanoparticles are prepared and characterized as described in ref [16]. These Fe$_3$O$_4$ nanoparticles are used as seed particles to grow Fe$_3$O$_4$ - CoFe$_2$O$_4$ core-shell particles. To a part of the Fe$_3$O$_4$ nanoparticle solution in trioctylamine (TOA), 0.16 mmol of Co-oleate and 0.27 mmol of Fe-oleate are added which are the starting precursors for the CoFe$_2$O$_4$ shell. The relative amounts of Fe and Co precursors necessary to obtain stoichiometric CoFe$_2$O$_4$ phase is optimized in control experiments where pure CoFe$_2$O$_4$ nanoparticles were prepared and analysed using ICP-OES (Inductively Coupled Plasma -  Optical Emission Spectroscopy). The reaction mixture is purged with nitrogen gas and heated to 300 $^{\circ}$C with continuous nitrogen flow and stirring. The temperature is held at 300 $^{\circ}$C for 15 minutes to enable the formation of CoFe$_2$O$_4$ shell on the Fe$_3$O$_4$ core. The core-shell particles are then collected after purification. The purification process involves collecting TOA in acetone after dissolving and centrifugation. The particles are then dispersed in chloroform and re-precipitated in ethanol before centrifuging to get solid powder of nanoparticles sample. A part of this sample is pressed into a pellet by cold pressing. The pellet is then annealed under Ar gas flow at 450 $^{\circ}$C for 2 hrs to remove organics from the surface of the particles and to ensure good inter-particle connectivity. Annealing conditions were optimized to ensure the highest connectivity between different nanocrystals, leading to a higher density and mechanical strength of the pellet, and lowering the electrical resistance between individual nanocrystals, while keeping the inter-diffusion between Fe$_3$O$_4$ and  CoFe$_2$O$_4$ to a minimum. Annealing at any higher temperature resulted in a degradation of the well-defined core-shell structure as represented by drastic changes in magnetic and transport properties due to the interdiffusion between the two components.

Fe$_3$O$_4$ - CoFe$_2$O$_4$ core-shell particles are checked for phase purity using a Bruker D8 Advance x-ray diffractometer. A JEOL High Resolution Transmission Electron Microscope (HRTEM) is used for checking the particle morphology, size and size distribution, while the energy dispersive x-ray spectroscopy (EDS) is used to confirm the existence of Co in several individual particles, indicating the formation of core-shell particles and not separate particles of Fe$_3$O$_4$ and CoFe$_2$O$_4$. Magnetic and magneto-transport properties were measured using Cryogenic and Quantum design PPMS systems. The magnetic properties are measured on both the as obtained powders and on the annealed pellets. The pellets are used for measuring magneto-transport properties.

\par Fe$_3$O$_4$ - CoFe$_2$O$_4$ core-shell nanoparticles form in a cubic spinel phase as confirmed from x-ray diffraction pattern (Fig. 1(a)). The HRTEM images in Figures 1(b) and 1(c) show nearly spherical particles of FO-CFO. Figure 1(d) shows the high-resolution images of individual particle with highly ordered crystal planes. The particle size distribution obtained from analysis of several TEM images is presented in Figure 1(e). This histogram is well described by the lognormal distribution function, shown by a continuous curve in the figure. This analysis establishes a diameter of 11.4 nm for FO-CFO which is distinctly larger than 10.2 nm, the diameter of Fe$_3$O$_4$ seed nanoparticles\cite{Anil20127} suggesting the growth of a few layers of CFO shell over the core FO particles. 
\par Careful EDS spectra collected on several individual particles of FO-CFO sample using HRTEM instrument looks similar to the representative spectrum presented in Figure 1(e). Each of the examined nanoparticles contains Co in addition to Fe, providing yet another evidence for the formation of core-shell particles. A simple calculation using the mean diameters of the bare\cite{Anil20127} and core-shell nanoparticles from TEM gives a value of 9.6 atomic percent for Co, which is close to the values of 8 to 9 atomic percent of Co estimated from the EDS analyses. However, the contrast difference between Co and Fe is not strong enough and the spatial resolution of the EDS not high enough for one to directly probe the nature of the FO-CFO interface in these core-shell nanoparticles. 
\begin{figure}
\includegraphics{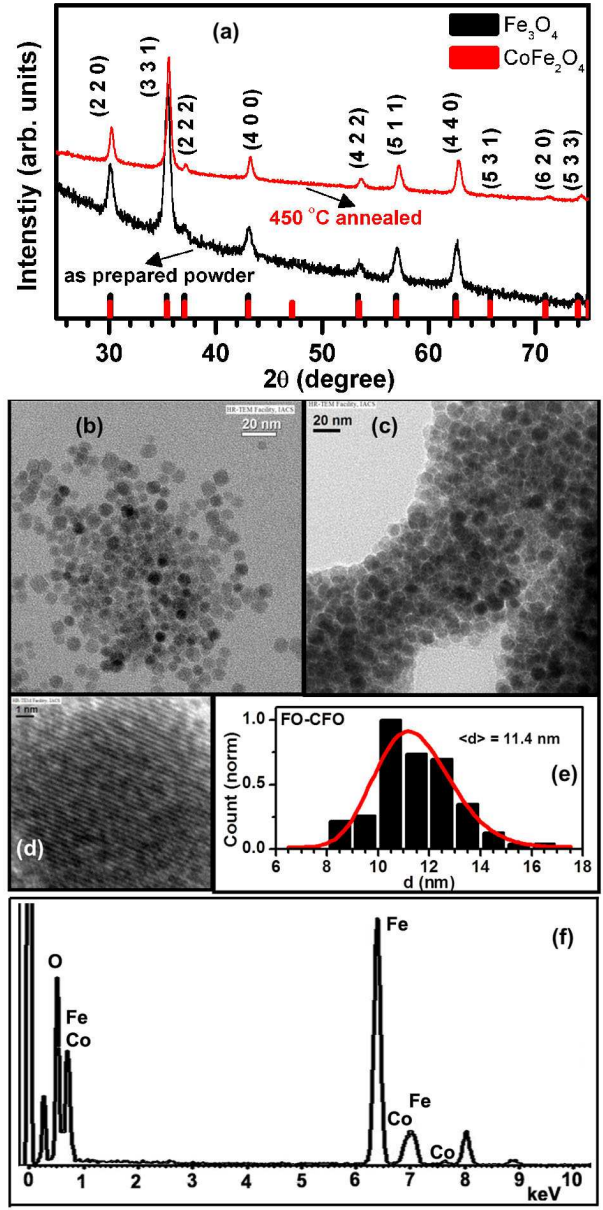}
\caption{\label{Fig.1} (Color online) \textbf{(a)} The XRD pattern of the core-shell Fe$_3$O$_4$ - CoFe$_2$O$_4$ (FO-CFO) is presented. The JCPDF data (black and red bars respectively for Fe$_3$O$_4$ and CoFe$_2$O$_4$) are also presented for comparison. The TEM images of the Fe$_3$O$_4$ - CoFe$_2$O$_4$ (FO-CFO) sample are shown in panels \textbf{(b)} and \textbf{(c)}, the panel \textbf{(d)} shows the high-resolution image of individual particle. The panel \textbf{(e)} shows the particle size distribution of the sample along with a lognormal curve fitting. The representative EDS data of the FO-CFO particle is given in panel \textbf{(f)}}
\vspace{-5pt}
\end{figure}

\par From the zero-field-cooled (ZFC) and field-cooled (FC) magnetization measurements shown in Figure 2(a), we estimate the blocking temperature, $T_{B}$, of the FO-CFO particles to be $\sim$ 220 K. It is to be noted that we do not observe any signature of a second $T_{B}$ which should have existed\cite{Masala81015} if there were also independent FO and/or CFO particles along with the FO-CFO core-shell particles. It should also be noted that CoFe$_2$O$_4$ shell of $\sim$ 0.6 nm thickness (from TEM analysis) equals to $\sim$ 28\% of the total volume of the particle of diameter $\sim$ 11.4 nm. This and the fact that the coercivity of the Fe$_3$O$_4$ and CoFe$_2$O$_4$ differ hugely results in the observed $M$($H$) loops where only CoFe$_2$O$_4$ coercivity is prominent as shown in Figure 2(b).
\par Figure 2(b) presents the $M$($H$) loops, at different temperatures, of the annealed pellet of FO-CFO and the representative $M$($H$) loops of the core-shell particles, before annealing, at two temperatures are shown in the inset to Figure 2(b). Results in Figure 2(b) clearly show that $M$($H$) loops measured on the pellet and powder forms have largely different coercivity values, with the pellets of FO-CFO exhibiting a huge increase in coercivity compared to the nanoparticle powder. For example, the coercivity at 50 K, of FO-CFO increases from 800 Oe in powder form to $\sim$ 12.3 kOe in the pellet form. This enhancement in the coercivity can be attributed to the increase in the effective crystallite size of CoFe$_2$O$_4$ in FO-CFO due to the annealing of pellet at 450 $^{\circ}$C, since annealing of the pressed pellet helps to fuse the individual shells of CFO on neighboring nanocrystals, thereby giving rise to a continuous matrix of CFO with FO embedded in it, increasing both the coercivity of the CFO component and the mechanical strength of the pellet as a whole. Since the coercivity is clearly a pronounced function of the CFO grain size and the growth in the grain size is a thermally driven statistical process, it is clear that the moderately sintered sample has a range of grain sizes, leading to a distribution of magnetic coercivities in the sample. The $M$($H$) data in the main frame of Figure 2(b) represents a weighted average of contributions from different grain sizes in the sample. \par In Figure 3, the magnetoresistance (\textit{MR}), defined as 100 $\times$ [($R$($H$)-$R(0)$)/$R(0)$], of FO-CFO pellet is plotted as a function of magnetic field ($H$) at 50 K. A peak in the \textit{MR}($H$) curve represents the highest resistive state of the sample. We represent the field value at which this resistance peak occurs by $H_{C}^{MR}$, in this manuscript.
\begin{figure}
\includegraphics{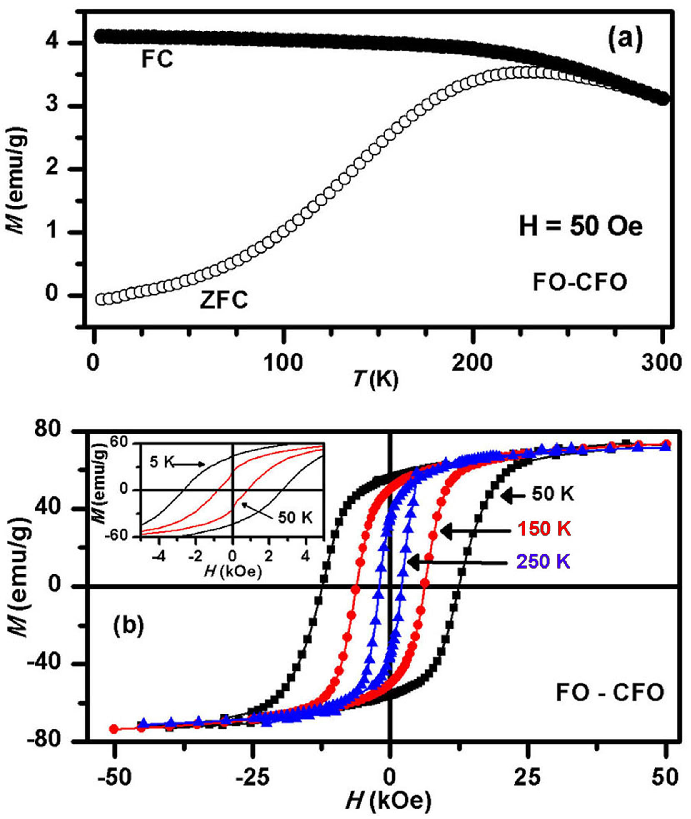}
\caption{\label{Fig.2 } (Color online) \textbf{(a)} The zero-field-cooled (ZFC) and field-cooled (FC) $M$($T$) data of as prepared Fe$_3$O$_4$ - CoFe$_2$O$_4$ (FO-CFO) particles, the cooling and measuring magnetic field is set as 50 Oe. The panel \textbf{(b)} shows the  hysteresis [$M$($H$)] loops of the annealed pellet of FO-CFO measured at different temperatures. The inset shows the hysteresis [$M$($H$)] loops measured on as prepared FO-CFO particles for two different temperatures, 5 and 50 K.}
\vspace{-10pt}
\end{figure}

\begin{figure}
\includegraphics{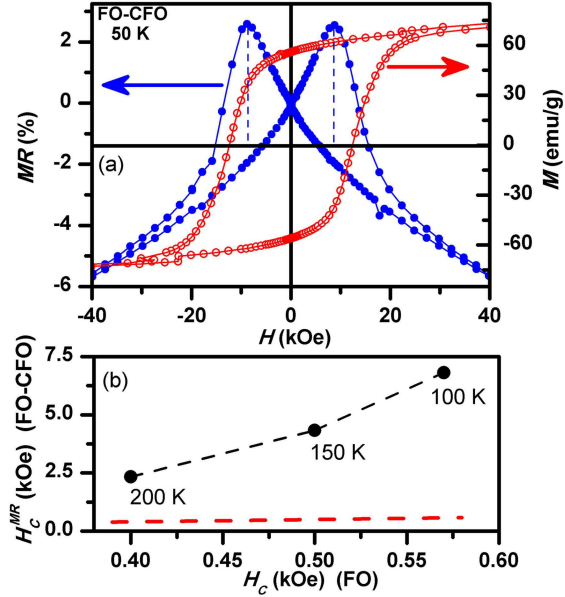}
\caption{\label{Fig.3 } (Color online) Representative plots, measured at 50 K, of magnetoresistance, (\textit{MR}) and magnetization, $M$ as a function of applied magnetic field, ($H$) of FO-CFO are shown in the panel \textbf{(a)}. In panel \textbf{(b)}, $H_{C}^{MR}$ values of FO-CFO sample are plotted as a function of $H_{C}$ of Fe$_3$O$_4$, for different temperatures. Since the scales for the two axes in Fig. 3(b) differ by more than an order of magnitude, we provide a dashed line representing $H_{C}^{MR}$ = $H_{C}$ of FO as a guide. This line shows the expected behavior of $H_{C}^{MR}$, if the tunnel barrier would be non-magnetic, thereby emphasizing the more than an order of magnitude enhancement of $H_{C}^{MR}$ on making the barrier magnetic.}
\end{figure}
As mentioned in the introduction, one of the characteristics of conventional TMR structures is that the $H_{C}^{MR}$ value of a TMR system, consisting of only one magnetic component, generally coincides with the magnetic coercivity of the metallic magnetic grains.\cite{Coey803815,Li711124} This is also true for the TMR of bare Fe$_3$O$_4$ nanoparticles in absence of any CFO shell covering it.\cite{Anil20127} In presence of the CFO over layer, however, the $H_{C}^{MR}$ observed for tunneling between Fe$_3$O$_4$ nanoparticles across the insulating CoFe$_2$O$_4$ barrier layer, is $\sim$ 9 kOe at 50 K (see Fig. 3(a)) which is much higher ( $\sim$ 13 times) than the $H_{C}$ value of 700 Oe, of Fe$_3$O$_4$ at 50 K. We have observed similarly large discrepancy between $H_{C}^{MR}$ of FO-CFO pellet compared to the $H_{C}$ of Fe$_3$O$_4$ nanoparticles at other temperatures as well. This is illustrated in Figure 3(b) where $H_{C}^{MR}$ of FO-CFO pellet has been plotted as a function of the $H_{C}$ of Fe$_3$O$_4$ annealed pellet (extracted from Ref [16]) for three different temperatures; the dashed line represents $H_{C}^{MR}$ = $H_{C}$ line, clearly showing more than an order of magnitude enhancement of $H_{C}^{MR}$ compared to the coercive field of Fe$_3$O$_4$ grains over the entire temperature range. Having established the irrelevance of the magnetic coercivity of the Fe$_3$O$_4$ nanoparticles in determining the $H_{C}^{MR}$ of the tunneling magnetoresistance between Fe$_3$O$_4$ nanoparticles separated by highly insulating CoFe$_2$O$_4$, we now show that $H_{C}^{MR}$ of the system is, in fact, controlled by the magnetic coercivity of the barrier layer, as proposed for the spin-valve type magnetoresistance in the case of SFMO,\cite{Sarma98157205,Ray9447007} in contrast to the usual TMR behavior. This is illustrated in Figure 4, by plotting $H_{C}^{MR}$ and $H_{C}$ of the sample as a function of the temperature. It is to be noted that the coercive fields of Fe$_3$O$_4$ and CoFe$_2$O$_4$ are so vastly different that the coercive field of the sample, seen in Figure 2, is essentially controlled entirely by the CoFe$_2$O$_4$ component of the sample. It is clear from Figure 4 that $H_{C}^{MR}$ is similar to $H_{C}$ at all temperatures, evidencing the control of the highest resistive peak in the \textit{MR} of the sample by the coercive field of the barrier layer, and not by the coercive field of the Fe$_3$O$_4$ grains, thereby establishing itself as an ideal example of a specifically engineered SVMR system. Here we note that the systematic (28-47\%) reduction of the $H_{C}^{MR}$ compared to $H_{C}$ is easy to understand. As already mentioned while discussing $M$($H$) plots of the sample in Figure 2, the sample is expected to be magnetically inhomogeneous to some extent, depending on the growth of the effective grain size of CoFe$_2$O$_4$ at various locations in the sample on annealing. The tunneling conductivity will be dominated by the thinnest CFO barrier layer separating the Fe$_3$O$_4$ grains and the coercivity of such thin parts of the CFO layer is indeed expected to be lower than the coercivity of the entire sample, averaged over all layer sizes. Thus, the magnetoresistance of the sample will reflect a somewhat lower coercivity than the average value, as indeed shown in Figure 4. It is instructive at this stage to compare the behavior of \textit{MR} in a system with Fe$_3$O$_4$ nanoparticles alone \cite{Anil20127} and in the present system using Fe$_3$O$_4$ - CoFe$_2$O$_4$ core-shell nanostructured material. It has been shown \cite{Anil20127} that \textit{MR} in the Fe$_3$O$_4$ system can be well understood in terms of the expected behavior of a traditional TMR material. It is important to note here that the \textit{MR} value of the present FO-CFO sample remains almost same as that for the Fe$_3$O$_4$ sample at comparable magnetic fields and identical temperatures.
\begin{figure}
\includegraphics{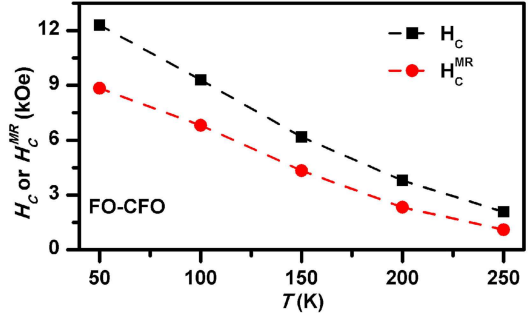}
\caption{\label{Fig.4} (Color online) (a) A comparison of magnetic coercivity, $H_{C}$ and the magnetoresistance coercivity, $H_{C}^{MR}$ of FO-CFO for different measurement temperatures is presented.}
\vspace{-10pt}
\end{figure}
 On the other hand, clearly we are able to influence the peak position of \textit{MR}($H$) in the core-shell system to make it larger than in the Fe$_3$O$_4$ system by more than a factor of ten. This clearly establishes that an artificial SVMR system can be engineered with controllable magnetic properties of the tunnel barrier.  In addition, we would also like to point out another difference in the \textit{MR}($H$) curves of Fe$_3$O$_4$ and FO-CFO samples. From Figure 3(a), it can be observed that the drop in resistance (or in \textit{MR}), after the peak, is sharp in the case of FO-CFO compared to the behavior of \textit{MR}($H$) for Fe$_3$O$_4$ alone. This sharper response in \textit{MR} in the case of FO-CFO system is consistent with the idea that the tunneling current in this system is controlled by the spin-valve type mechanism and can be understood in the following terms. For $H$ $<$  $H_{C}^{MR}$ even though the magnetic field is sufficient to saturate the FO core moment\cite{Anil20127}, the tunneling current is blocked until the tunnel barrier magnetically aligns with the FO core and for $H$ $>$ $H_{C}^{MR}$ the tunneling current is allowed suddenly, leading to a sharp fall in resistance (or in \textit{MR}). It should also be noted that the \textit{MR} results observed in FO-CFO are distinctly different from the \textit{MR} results on Co doped Fe$_3$O$_4$ \cite{Tripathy101013904} where no sharp fall in resistance is observed and further a clear reduction in \textit{MR}\% compared to undoped Fe$_3$O$_4$ due to loss of spin-polarization with Co doping is also reported.

In conclusion, magnetoresistance properties of a tunneling magnetoresistance system with a magnetic tunnel barrier are studied by using a specifically engineered core-shell nanoparticle structure, Fe$_3$O$_4$-CoFe$_2$O$_4$,  as a model system. It is observed that the highest resistive state of such a system is essentially controlled by the magnetic coercivity of the tunnel barrier in contrast to the standard TMR behavior and defining a spin-valve type MR system. In addition, the magnetoresistance value of the core-shell nanoparticle system is comparable to the bare Fe$_3$O$_4$ nanoparticle system indicating that the spin polarization for Fe$_3$O$_4$ core is not diminished in the present case unlike in Co doped Fe$_3$O$_4$ samples. Thus, our results provide an alternate route to manipulate the magnetoresistance behavior of a TMR system by using spin-valve like action of the hard magnetic barrier layer without any quantitative degradation of the MR.
\vspace{15pt}

Authors thank Department of Science and Technology for supporting this research. PAK thanks the Swedish Research Council (VR) and the G\"{o}ran Gustafsson Foundation, Sweden for funding. SC acknowledge support by the Japan Society for the Promotion of Science (JSPS) through the ``Funding Program for World-Leading Innovative R\&D on Science and Technology (FIRST Program)".


%

\end{document}